\documentclass[preprint]{aastex}

\shorttitle{PALFA intermittent pulsars}
\shortauthors{A. G. Lyne et al.}

\begin{document}

\title{Two long-term intermittent pulsars discovered in the PALFA Survey}

\author{
A.~G.~Lyne\altaffilmark{1}, 
B.~W.~Stappers\altaffilmark{1},
P.~C.~C.~Freire\altaffilmark{2},
J.~W.~T.~Hessels\altaffilmark{3,4},
V.~M.~Kaspi\altaffilmark{5},
B.~Allen\altaffilmark{6,7,8},
S.~Bogdanov\altaffilmark{9},
A.~Brazier\altaffilmark{10},
F.~Camilo\altaffilmark{11},
F.~Cardoso\altaffilmark{7},
S.~Chatterjee\altaffilmark{10},
J.~M.~Cordes\altaffilmark{10},
F.~Crawford\altaffilmark{12},
J.~S.~Deneva\altaffilmark{13},
R.~D.~Ferdman\altaffilmark{5},
F.~A.~Jenet\altaffilmark{14}, 
B.~Knispel\altaffilmark{6,7},  
P.~Lazarus\altaffilmark{2}, 
J.~van~Leeuwen\altaffilmark{3,4}, 
R.~Lynch\altaffilmark{5}, 
E.~Madsen\altaffilmark{5},
M.~A.~McLaughlin\altaffilmark{15},
E.~Parent\altaffilmark{5},
C.~Patel\altaffilmark{5},
S.~M.~Ransom\altaffilmark{16},
P.~Scholz\altaffilmark{5},
A.~Seymour\altaffilmark{17},
X.~Siemens\altaffilmark{7}, 
L.~G.~Spitler\altaffilmark{2},
I.~H.~Stairs\altaffilmark{18,5}, 
K.~Stovall\altaffilmark{19,20}, 
J.~Swiggum\altaffilmark{7},
R.~S.~Wharton\altaffilmark{10},
W.~W.~Zhu\altaffilmark{2}
}

\altaffiltext{1}{Jodrell Bank Centre for Astrophysics, School of Physics and Astronomy, Univ. of Manchester, Manchester, M13 9PL, UK} 
\altaffiltext{2}{Max-Planck-Institut f\"ur Radioastronomie, Auf dem H\"ugel 69, D-53121 Bonn, Germany} 
\altaffiltext{3}{ASTRON, The Netherlands Institute for Radio Astronomy, Postbus 2, 7990 AA, Dwingeloo, The Netherlands} 
\altaffiltext{4}{Anton Pannekoek Institute for Astronomy, Univ. of Amsterdam, Science Park 904, 1098 XH Amsterdam, The Netherlands}
\altaffiltext{5}{Dept.~of Physics and McGill Space Institute, McGill Univ., Montreal, QC H3A 2T8, Canada}
\altaffiltext{6}{Max-Planck-Institut f\"ur Gravitationsphysik, D-30167 Hannover, Germany}
\altaffiltext{7}{Physics Dept., Univ. of Wisconsin - Milwaukee, 3135 N. Maryland Ave., Milwaukee, WI 53211, USA}
\altaffiltext{8}{Leibniz Universit{\"a}t Hannover, D-30167 Hannover, Germany}
\altaffiltext{9}{Columbia Astrophysics Laboratory, Columbia Univ., New York, NY 10027, USA} 
\altaffiltext{10}{Dept. of Astronomy, Cornell Univ., Ithaca, NY 14853, USA} 
\altaffiltext{11}{SKA South Africa, Pinelands, 7405, South Africa} 
\altaffiltext{12}{Dept. of Physics and Astronomy, Franklin and Marshall College, Lancaster, PA 17604-3003, USA} 
\altaffiltext{13}{National Research Council, resident at the Naval Research Laboratory, Washington, DC 20375, USA}
\altaffiltext{14}{Center for Gravitational Wave Astronomy, Univ. of Texas - Brownsville, TX 78520, USA} 
\altaffiltext{15}{Dept. of Physics, West Virginia Univ., Morgantown, WV 26506, USA} 
\altaffiltext{16}{NRAO, Charlottesville, VA 22903, USA} 
\altaffiltext{17}{Arecibo Observatory, HC3 Box 53995, Arecibo, PR 00612, USA} 
\altaffiltext{18}{Dept.~of Physics and Astronomy, Univ.~of British Columbia, Vancouver, BC V6T 1Z1, Canada} 
\altaffiltext{19}{NRAO, PO Box 0, Socorro, NM 87801, USA}
\altaffiltext{20}{Dept. of Physics and Astronomy, Univ.~of New Mexico, NM 87131, USA} 

\begin{abstract}

We report the discovery of two long-term intermittent radio pulsars in
the ongoing Pulsar Arecibo L-band Feed Array survey.  Following discovery with the
Arecibo Telescope, extended observations of these pulsars over several
years at Jodrell Bank Observatory have revealed the details of their
rotation and radiation properties.  PSRs J1910+0517 and J1929+1357
show long-term extreme bimodal intermittency, switching between
active (ON) and inactive (OFF) emission states and indicating the
presence of a large, hitherto unrecognized, underlying population of
such objects.  For PSR~J1929+1357, the initial duty cycle was
$f_{\rm ON}$=0.008, but two years later, this changed quite abruptly to
$f_{\rm ON}$=0.16.  This is the first time that a significant evolution in
the activity of an intermittent pulsar has been seen, and we show that
the spin-down rate of the pulsar is proportional to the activity. The
spin-down rate of PSR~J1929+1357 is increased by a factor of 1.8 when
it is in active mode, similar to the increase seen in the other three
known long-term intermittent pulsars.  These discoveries increase the
number of known pulsars displaying long-term intermittency to
five. These five objects display a remarkably narrow range of
spin-down power ($\dot{E}\, \sim \,10^{32}\, \rm erg \, s^{-1}$) and
accelerating potential above their polar caps. If confirmed by further
discoveries, this trend might be important for understanding the
physical mechanisms that cause intermittency.

\end{abstract}

\section*{}\eject  % force text to start in second column

\section{Introduction}\label{sec:intro}

Intermittent pulsars offer a unique opportunity to study the
relationship between the spin-down and the emission of radio pulsars
\citep{klo+06}.  These pulsars show normal pulsar emission properties
for a period of time (ON phase) and then switch OFF and back ON again,
with cycle times measured in days or even years.  While many pulsars
exhibit such switching behavior on timescales of seconds to fractions
of a day (a phenomenon generally known as ``pulse nulling''), the
long cycle times present the possibility of determining the rotational
slow-down rates in both the ON and OFF states.  Only three such
objects are known, but they offer a rare opportunity to study the
effect of particle flows in pulsar magnetospheres on the spin-down
rates and hence the braking torques of these neutron stars
\citep{klo+06,crc+12,llm+12}.

A similar but somewhat less dramatic phenomenon is ``mode changing,''
in which switching occurs between two (or occasionally more) modes in
which the pulse profiles or flux densities are different
\citep{bac70,lyn71,msfb80,fmw81}, and sometimes occurs in X-rays as well as
in radio \citep{hhk+13,mkt+16}. Again, the timescales vary from pulsar to
pulsar from seconds to years, and furthermore the longest-cycle-time
objects have revealed that the two modes have different slow-down rates
\citep{lhk+10}, albeit with much smaller differences. The close
similarity of the nulling/intermittent and mode-changing phenomena has
led authors to suspect that they are closely related
\citep[e.g.][]{ls05a,wmj07}.

Dramatic changes in the emission and spin properties have
important implications for the present and long-term evolution of
these systems and perhaps for the pulsar population as a whole. There
have therefore been a number of mechanisms proposed for both those
objects where the radio emission completely switches OFF and
mode-changing sources. There is a range of models which consider that the
plasma supply to the magnetosphere affects the global charge
distribution \citep[e.g.][]{tim10,kkhc12,lst12} or that the plasma is
moving at different velocities \citep{mh14}. Other proposed mechanisms
consider influences closer to the neutron star surface, such as changes
in the properties of the particle acceleration region \citep{smg15} or
twists in the magnetic field structure \citep{hyt16}. The quasi-periodic
nature of the intermittency in some systems has led to the proposal
that it may be related to free precession \citep[e.g.][]{alw06,jon12}
or that there may be some sort of forcing mechanism
\citep[e.g.][]{cs08,rks+08,mbh13}. Others have suggested that it
is a chaotic \citep{sl13} or Markov \citep{cor13} process.

Because the sporadic nature of intermittent pulsars makes it very
difficult to discover them in one-pass surveys, they are also
representatives of a much larger underlying population \citep{klo+06}.
We note that such objects are similar to rotating radio transients
(RRATs) which are often difficult to detect and which also represent a
much larger underlying population.  The existence of such hidden
populations opens up the possibility that the birthrate of neutron
stars may prove to exceed the rate of supernova collapse events and
demand the presence of other formation processes.  It is clearly
important to increase the number of known intermittent pulsars.

In this paper, we present the discovery and subsequent study of two
extreme long-term intermittent pulsars, PSR~J1910+0517 and
PSR~J1929+1357, which were found as part of the the ongoing Pulsar
Arecibo L-Band Feed Array (PALFA) project.  This is a deep pulsar
survey of low Galactic latitudes being undertaken using the 305 m William
E.\ Gordon Telescope at the Arecibo Observatory.  The survey is
described and its parameters are discussed in \citet{cfl+06},
\citet{vcl+06}, \citet{dcm+09}, \citet{lab+12}, \citet{nab+13},
\citet{slm+14}, and \citet{lbh+15}.

Following discovery of the pulsars at Arecibo, the 76-m Lovell
Telescope at the Jodrell Bank Observatory, United Kingdom, has been
used within the PALFA collaboration both to confirm the existence of
and to conduct follow-up timing observations of, more than half of the
169 pulsars detected hitherto in the PALFA survey and to conduct
follow-up timing observations of these pulsars.  Having confirmed
the existence of the two pulsars, we, because of their potential
importance, we have made extensive further observation of these
sources with the Lovell Telescope, in order to further study their
emission and timing properties, and we report on these studies in this
paper. In \S\ref{sec:obs} we briefly describe the discovery and
follow-up observations, and in \S\ref{sec:intermittent} we discuss in
turn the two pulsars which display the extreme long-term
intermittency.  We discuss our conclusions in \S\ref{sec:discussion}.

\section{Observations}\label{sec:obs}

Since the PALFA survey is described thoroughly elsewhere (see
\S\ref{sec:intro}), we give here only a brief summary here of the survey
observations as described by \citet{lsb+16}.

The survey area covers the two regions close to the Galactic plane
($|b|<5^{\circ}$) which are observable using the Arecibo Telescope.
These are located at Galactic longitudes $32^{\circ} \lesssim \ell \lesssim 77
^{\circ}$ and $168{^\circ} \lesssim \ell \lesssim 214^{\circ}$.  The
survey utilizes the seven simultaneous independent dual-polarization
beams provided by the ALFA cryogenic receiver.  Data were collected
for 268\,s for each telescope pointing from 322 MHz passbands centered
on 1375\,MHz.  For the discovery of the two pulsars, the PALFA
observations used the Mock spectrometers to produce 960 frequency
channels across the passband of each polarization channel.  The
frequency channels are sampled with 16-bit precision every 65.5~$\mu$s
and stored on disk for off-line processing.  The ``Quicklook''
pipeline was used to discover both pulsars. This process analyzes the
data shortly after they are collected for dispersed periodic signals
\citep{cfl+06,lbh+15}.

Confirmation and all subsequent observations of the two pulsars
were all carried out with the 76 m Lovell Telescope at Jodrell Bank
Observatory using a dual-polarization cryogenic receiver which had a
cold-sky noise system equivalent flux density of 25 Jy.  A digital
filterbank was used to receive data in a passband of 1350 MHz to
1700 MHz with 0.5 MHz bandwidth channels.  For each polarization, the
power from the two channels was then folded and dedispersed at the
nominal period and dispersion measure of the pulsar.  The observations
at Jodrell Bank reported here were mostly made between 2011 November
and 2016 January (MJD~55900-57400).

\section{Intermittent pulsars: PSRs~J1910+0517 and J1929+1357}\label{sec:intermittent}

Shortly after the discovery, follow-up observations of PSRs
J1910+0517 and J1929+1357, follow-up observations indicated that both 
objects had a bimodal emission
nature and were frequently undetected, suggesting that they suffered
extreme long-term intermittency.  The large values of DM for these two
pulsars (300 and 151 cm$^{-3}$pc, respectively) and the bimodal nature
of their flux densities indicate that the intermittency is not due to
interstellar scintillation. 

These pulsars were therefore
subjected to intensive monitoring programs using the Lovell Telescope
to investigate these properties. The results of these observations are
discussed below.

\subsection{Intermittent Pulsar: PSR~J1910+0517}\label{sec:J1910+0517}

\begin{figure*}[p]
\begin{center}
\includegraphics[width=4.0in,angle=270]{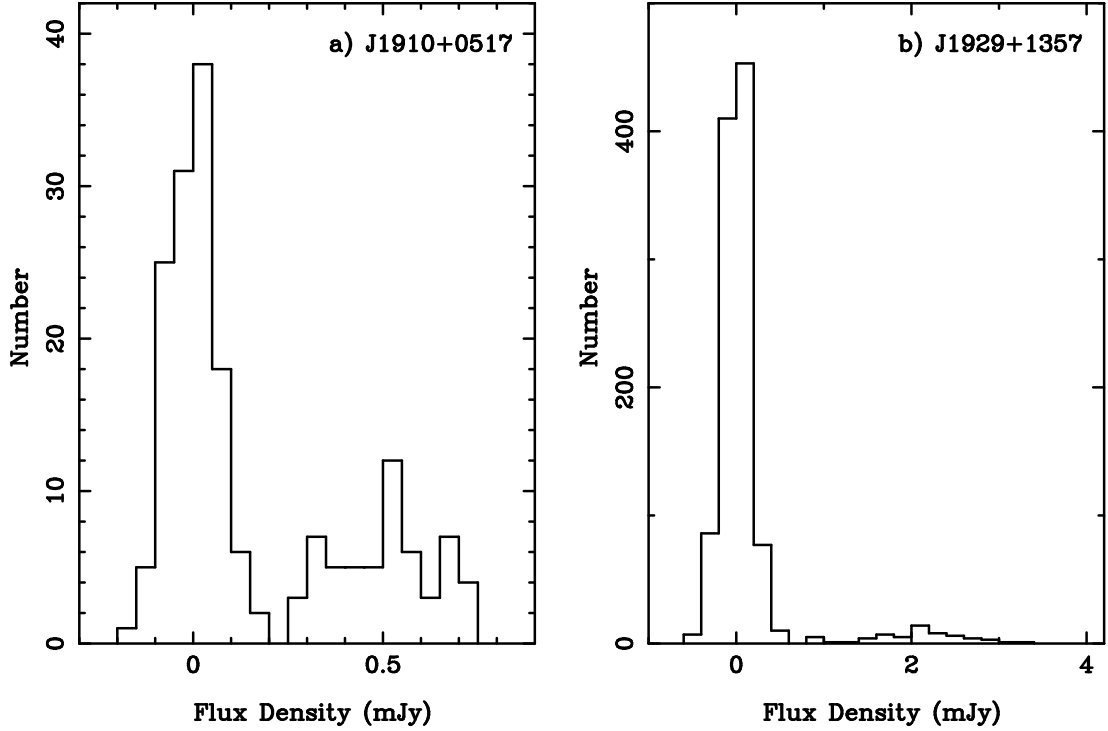}
\caption{\label{fig:hists} Histograms of the mean pulsed flux density
of (a) each of the 179 observations of PSR~J1910+0517 and (b) of the
1084 observations of PSR~J1929+1357.  Six observations of
PSR~J1910+0517 and for 16 of the observations of PSR~J1929+1357, in
which the pulsar was emitting for only a fraction $f$ of the
observation duration, the observations were split into two at the
switch between the ON and OFF
phases and two corresponding flux densities were calculated.  
The widths of the peaks centered on flux density
of zero are determined primarily by the statistical uncertainty in the
flux density measurements.  Note the clear bimodal nature of the
distributions, which indicates that the pulsars are emitting on
average for only about 30\% and 5\% of the time, respectively.  The
widths of the non-zero peaks may also reflect any intrinsic
fluctuations in the pulsar emission, but any interstellar diffraction
scintillation will mostly have been averaged out across the large receiver
bandwidth.}
\end{center}
\end{figure*}

PSR~J1910+0517 was first detected at Arecibo in an observation made on
2011 November 10 (MJD 55875) and was confirmed at Jodrell Bank later that
month, on 2011 November 29 (MJD 55894).  A total of 179 observations
were subsequently made of this pulsar up to 2016 January 13 (MJD 57400).
The observations had durations between 20 and 30 minutes and were
cleaned of significant radio-frequency interference.  The pulsar was
clearly detected 57 times and was undetected on the other 122 occasions.
The distribution of the observed mean pulsed flux density in these
observations is summarized in Figure \ref{fig:hists}a and shows the
marked bimodal nature, with one portion of observations centered on
zero, in which pulsed radio emission is undetectable (the OFF phase).
The second, separate and wider distribution is centered on a mean of
0.51~mJy, with standard deviation of 0.13~mJy (the ON phase).
However, a careful inspection of the 3 minute subintegrations of each
observation reveals that the pulsar clearly switched emission states
during 6 of the 179 observations.  During the total observation time
of 67.1 hr, the pulsar was ON for 20.1 hr and the duty cycle, the
fraction of time spent in the ON state, is $f_{\rm ON}=0.30(4)$.

\begin{figure*}[p]
\begin{center}
\includegraphics[width=4.0in,angle=0]{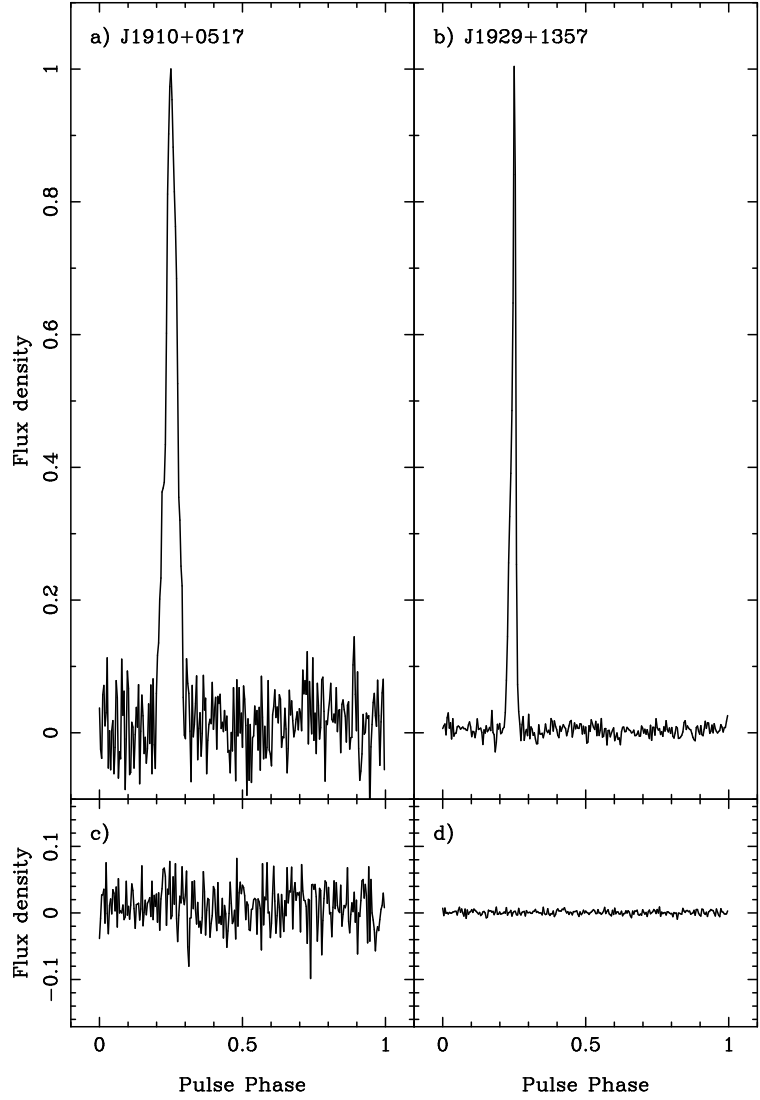}
\caption{\label{fig:profs} Integrated pulse profiles for
PSRs~J1910+0517 and J1929+1357 at 1520 MHz. (a) and (b) are the ON
profiles for the
two pulsars, respectively, scaled to a peak flux density of 1, while (c)
and (d) are the two OFF profiles, presented on the same scale as (a) and
(b). For PSR~J1910+0517, the integration times for the ON and OFF
profiles were 20 and 47 hr, while for PSR~J1929+1357, the integration
times were 8 and 150 hr.}
\end{center}
\end{figure*}

The integrated profiles of all the ON observations and all the OFF
observations were formed after alignment using the ephemeris given in
Table~\ref{table:intermittent} and are presented in
Figure~\ref{fig:profs}a and \ref{fig:profs}c.  The half-power width of
the pulse, $W_{50}$=13.0~ms, is unremarkable.  An inspection of the OFF
emission-mode profile suggests the possibility of some low-level
emission at the longitude of the ON pulse.  This amounts to a fraction
of 0.02(1) of the ON pulse emission. This may be due either to the 
misidentification of some ON-state subintegrations as being OFF due to
signal-to-noise limitations or to the OFF emission state not being a
pure ``null'' state but one with a small amount of emission.  Other examples
of pulsars with low-level emission modes are PSRs~B0826$-$34
\citep{elg+05} and J1853+0505 \citep{yws+15}.

\begin{figure*}[p]
\begin{center}
\includegraphics[width=5.0in,angle=0]{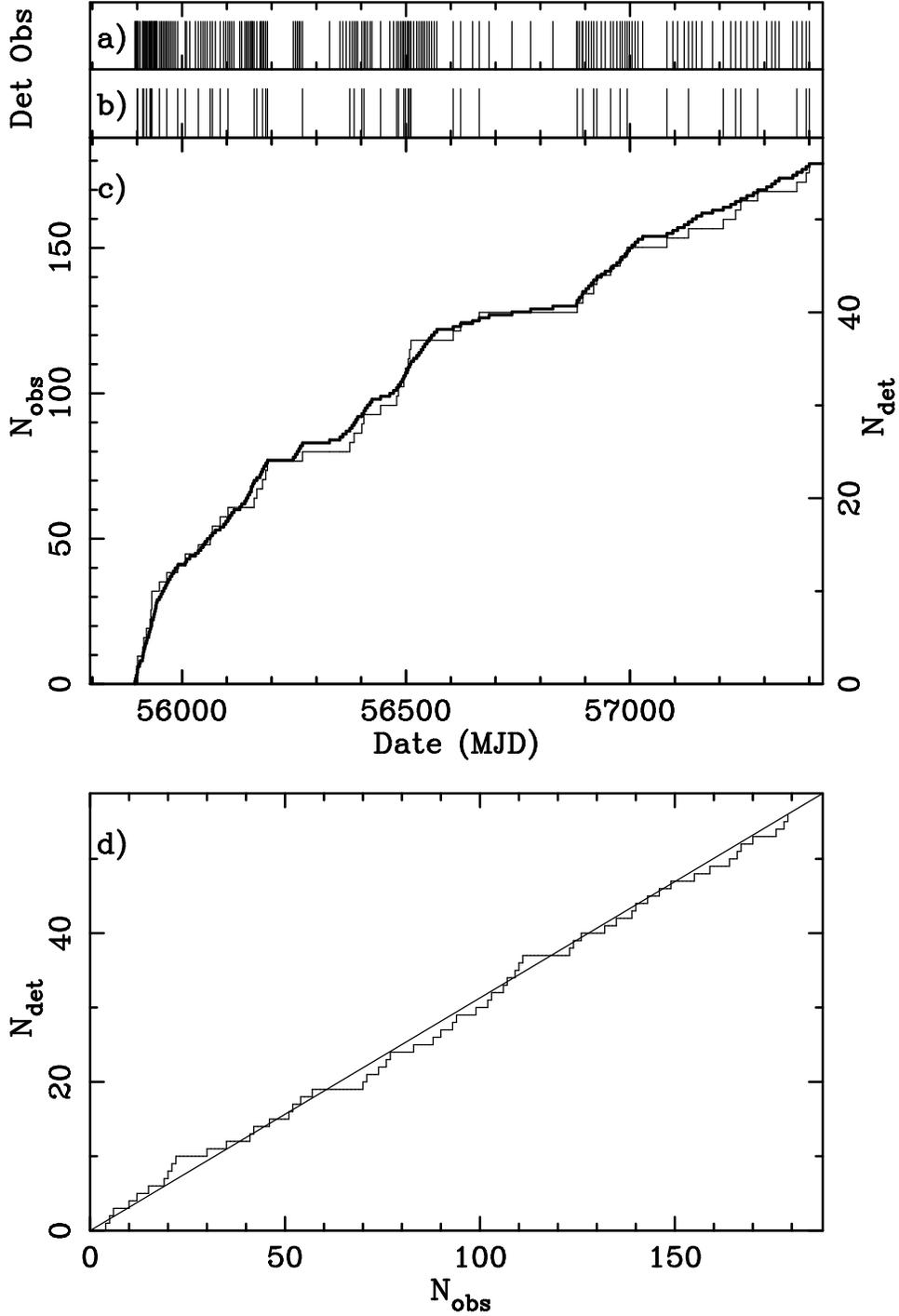}
\caption{\label{fig:1910_stats} The detection history of
PSR~J1910+0517. (a) The MJDs of observation of the pulsar. (b) The MJDs
of the observations in which the pulsar was detected, representing
about one third of the total observations. (c) Cumulative
plots of the numbers of detections (N$_{\rm det}$, thin line) and the
observations (N$_{\rm obs}$, heavy line). The similar forms of these two
curves suggest that the fraction of observations when the pulsar is
active is unchanging.  (d) Cumulative plot of the number of detections
plotted against the observation number of the 179 observations of
PSR~J1910+0517. The straight line has a slope which is equal to the
mean duty cycle $f_{\rm ON}$ of the pulsar, 0.31(4) over the 4 years of
observation. The local slopes are all consistent with this duty cycle.
}
\end{center}
\end{figure*}

The distributions of the observations and the detections over the 4 yr
period are summarized in Figures~\ref{fig:1910_stats}a-c. It is notable
that the number of detections of the pulsar closely tracks the total
number of observations made of the source.  This is more directly seen
in Figure~\ref{fig:1910_stats}d, in which the cumulative number of
detections is plotted against the cumulative number of observations.
The local slope in such a diagram represents the duty cycle $f_{\rm
ON}$, the fraction of the time spent in the ON emission state.  The
straight line corresponds to a mean value of $f_{\rm ON}$=0.30(4). The
absence of any systematic deviation of the data from this line
suggests that this value is unchanging over the 4 year period.

As indicated above, most of the observations are either ON or OFF for
the whole of their 20-30 min duration, indicating that the typical
timescale between switches of state is substantially greater than
this, and it would require a large investment of telescope time to
track the switches directly. However, we can estimate the timescale
statistically from the 6 changes of state seen in the 67.1 hr 
total duration of the 179 observations.  Since there are two changes
of state for each ON/OFF cycle, the average cycle time is $t_{\rm
CYCLE}$=22(9) hr.  The average ON, or active, time is therefore
$t_{\rm ON}=f_{\rm ON}\times t_{\rm CYCLE}$=7(3) hours and the average
time spent OFF is $t_{\rm OFF}$=15(6) hr.
 
\begin{figure*}[p]
\begin{center}
\includegraphics[width=4.0in,angle=270]{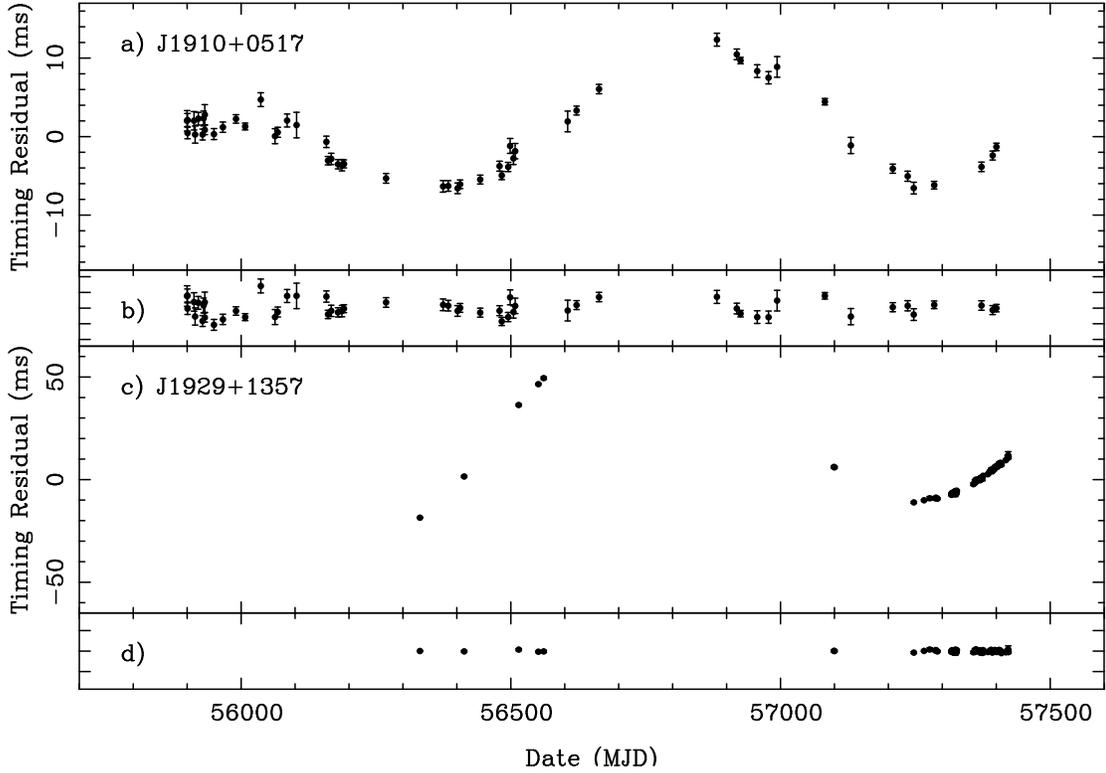}
\caption{\label{fig:resids}Timing residuals of the pulse arrival times
of PSRs J1910+0517 and J1929+1357 relative to simple spin-down models.
For each of the two pulsars, the residual plots in (a) and (c) were
made by performing timing fits for just the spin-period and spin-down
rate, with the best-fit positions given in Table
\ref{table:intermittent} held fixed, indicating the levels of timing
noise. For (b) and (d) the residual plots were made relative to the full fits
given in Table~\ref{table:intermittent} which include four and three
frequency derivatives, respectively.}
\end{center}
\end{figure*}

Times of arrival (TOAs) were obtained for all the ON observations by
cross-correlation of the profiles with a standard template and
processed using standard analysis techniques with the {\sc
Psrtime}\footnote{\url{http://www.jb.man.ac.uk/pulsar/observing/progs/psrtime.html}}
and {\sc Tempo}\footnote{\url{http://tempo.sourceforge.net}} software
packages.  A coherent timing fit has been made of a standard slow-down
timing model to the TOAs.  This model included the pulsar position, the
rotation frequency, and its first derivative.  The timing residuals,
the difference between the TOAs and the fitted model, are shown in
Figure \ref{fig:resids}a, in which there is clearly significant timing
noise.  A satisfactory description of the TOAs requires a fit for the
frequency and its first four derivatives.  The parameters of this fit
are summarized in Table~\ref{table:intermittent}, together with the
statistics discussed above.

\subsection{Intermittent Pulsar: PSR~J1929+1357}\label{sec:J1929+1357}

PSR~J1929+1357 was first detected at Arecibo in an observation made on
2012 September 20 (MJD 56190).  The first attempt at confirmation was made
at Jodrell Bank at UT 03:50 on 2013 February 8 (MJD 56331), but there was
no evidence for any pulsations.  However, the pulsar was clearly seen in a
second observation at UT 14:04 on the same day, with a signal-to-noise
ratio of 53 in 17 minutes.

However, during the next nine months, up to 2013 November 05 (MJD 56601), a
total of 656 observations of either 6 minute or 12 minute duration were made
of this pulsar, in which it was detected just 5 times, with
signal-to-noise ratios of between 15 and 30, and was undetected on the
other 651 occasions. During this period of time, the mean duty cycle
was thus $f_{\rm ON}\sim0.008$, so that the 5 detections required
around 100 hr of telescope time. Apart from confirmation of the
existence of this pulsar and the determination that the duty cycle was
very small, no other astrophysical information was gained during this
time, mainly because the sparseness of the detections
prevented the establishment of a coherent timing solution. The use of
further telescope time could not be justified, and observations were
discontinued.

After a break of nearly 15 months, observations of the pulsar were
restarted at Jodrell Bank in 2015 January in order to check that
there was no change in this behavior.  Indeed, spread over the next
six months, a further 90 observations, amounting to 12 hr of
telescope time, yielded two more detections, representing a barely
significant increase in the duty cycle. However, in August and
September of that year, a further 40 observations yielded 7
detections, indicating a significant uplift in the detection rate. The
cadence of observation was increased, and by 2016 mid-January (MJD
57407), a further 47 detections were made from just 317 observations,
making a total of 61 detections from 1084 observations.

The distribution of the observed mean pulsed flux density in all these
observations is summarized in Figure \ref{fig:hists}b, 
clearly showing a bimodal nature, with one portion of observations centered
on 0.002(5)~mJy and having a standard deviation of 0.15~mJy about the
mean.  The second, separate and wider distribution is centered on a
mean of 2.2~mJy, with standard deviation of 0.6~mJy, and represents
about 6\% of the observations.  This flux density is more than 10
times the median flux density of the 64 long-period pulsars discovered
so far in the PALFA survey \citep{nab+13,lsb+16}, making this pulsar one of the
brightest PALFA pulsars.

Integrated profiles of all the ON observations and all the OFF
observations were formed and are presented in Figures~\ref{fig:profs}b
and \ref{fig:profs}d.  The half-power width of the pulse, of
$W_{50}$=11.5~ms, is unremarkable.

\begin{figure*}[p]
\begin{center}
\includegraphics[width=5.0in,angle=0]{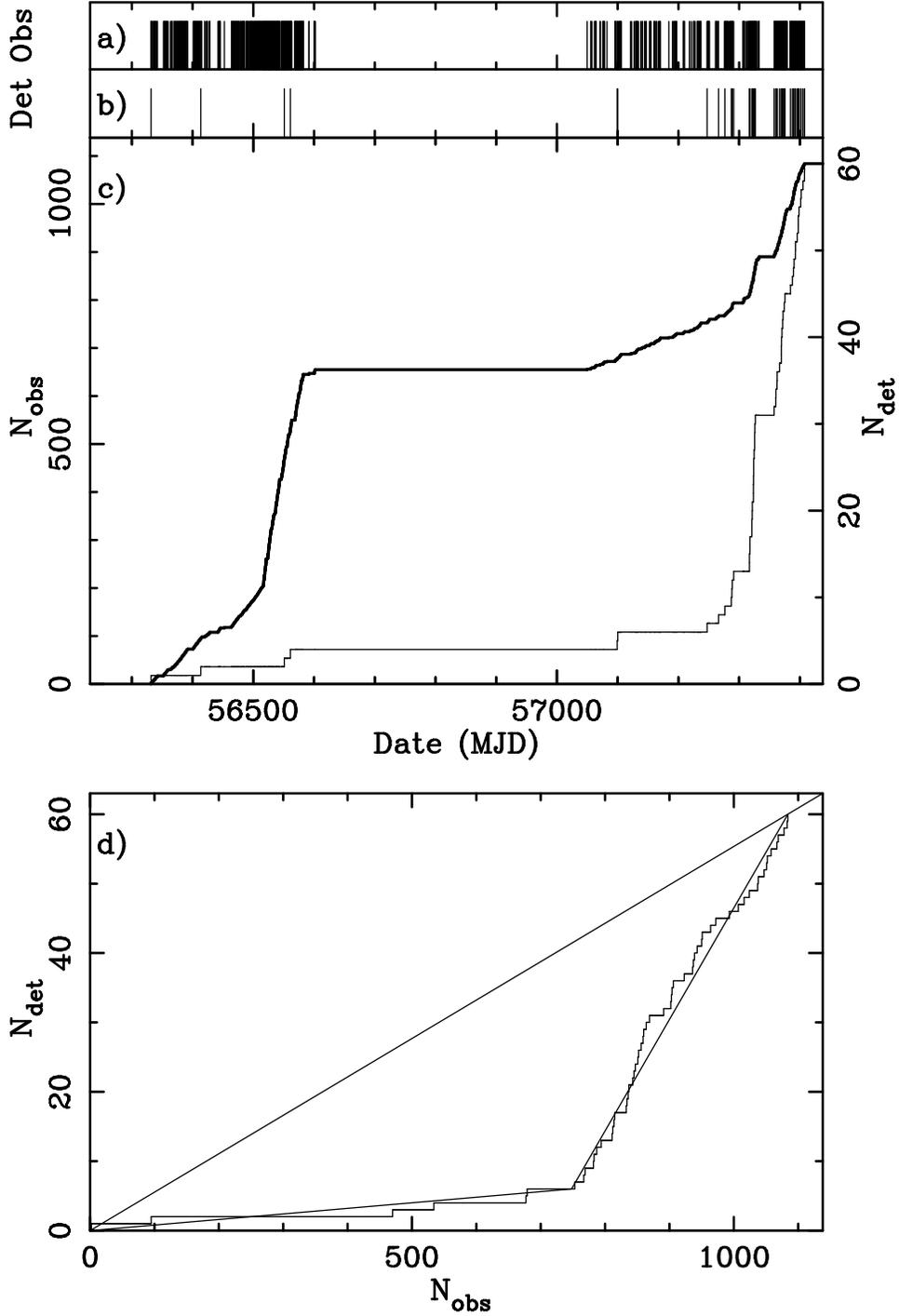}
\caption{\label{fig:1929_stats} The detection history of
PSR~J1929+1357. (a) The MJDs of the 1084 observations of the pulsar. (b)
The MJDs of the 61 positive detections of the pulsar, representing a
small fraction of the observations in (a).  (c) The cumulative plots of
the numbers of detections (N$_{\rm det}$, thin line) and the
observations (N$_{\rm obs}$, heavy line). The very different forms of
these two curves suggest that the fraction of observations when the
pulsar is active has changed significantly during the experiment.  (d)
Cumulative plot of the number of detections N$_{\rm det}$ plotted
against the observation number N$_{\rm obs}$, of the 1084 observations of
PSR~J1929+1357. The straight line has a slope which is equal to the
mean duty cycle $f_{\rm ON}$ = 0.055(7) over the 3 years of observation.
The data are described approximately by two straight-line portions
with a break at around 2015 Aug 1 (MJD~57235).  These lines represent
mean duty cycles of $f_{\rm ON}$ = 0.009(4) and $f_{\rm ON}$ = 0.16(2)
respectively.}
\end{center}
\end{figure*}

The distribution of the observations and the detections over the 3 yr
period are summarized in Figures~\ref{fig:1929_stats}a-c. Unlike that
of PSR~J1910+0517 described above, the rate of detection of this
pulsar does not track the rate of observations of the source.  This is
more directly seen in Figure \ref{fig:1929_stats}d, in which the local
value of the slope represents the duty cycle $f_{\rm ON}$, the
fraction of the time spent in its ON emission state. The diagonal
straight line corresponds to a mean value of $f_{\rm
ON}$ = 0.055(7). However, the data are well described by two separate
approximately linear sections with a break around MJD~57235.  In 2013 
and for the first 6 months of observation in 2015 (MJD~56331-57235),
the mean duty cycle is $f_{\rm ON}$ = 0.009(4), while, following this
(MJD~57235-47408), the mean duty cycle has increased by over an order
of magnitude to $f_{\rm ON}$ = 0.16(2).

Most of the observations are either ON or OFF for their whole 6-12 minute
duration, indicating that the typical timescale between switches of
state is substantially greater than this, requiring a large investment
of telescope time to track the switches directly. However, as with
PSR~J1910+0517, we can estimate the timescale statistically from the
16 changes of state seen during the 157.3 hr total duration of the
1084 observations.  Since, on average, there are two changes of state
for each cycle, the average cycle time is $t_{\rm CYCLE}$ = 20(5) hr.
The average ON, or active, time is $t_{\rm ON}=f_{\rm ON}\times t_{\rm
CYCLE}$ = 1.1(3) hr.

Repeating these calculations separately for the two sections spanning
the break at MJD~57235, we get $t_{\rm CYCLE}$ = 220(110) and 6.3(1.6)
hr and $t_{\rm ON}$ = 2(1) and 1.0(3) hr respectively.

The increased detection rate toward the end of 2015 has allowed a
coherent timing fit to all the TOAs obtained for PSR~J1929+1357
since the start of 2013.  There is clearly significant timing noise,
and as can be seen in Figure \ref{fig:resids}b, the data for this
pulsar do not fit a simple spin-down model, and a satisfactory
ephemeris requires a fit for frequency and its first three
derivatives.

During recent years, several pulsars have been shown to display
changes in emission properties which are related to their spin-down
rate.  The changes in emission always seem to be related to sudden,
switched changes between two discrete states.  For several pulsars,
the changes in emission are seen as pulse shape changes
\citep{lhk+10,bkj+16}, but for three pulsars ---- B1931+24 \citep{klo+06},
J1832+0029 \citep{llm+12} and J1841$-$0500 \citep{crc+12} ---- are
seen to switch between emission as normal pulsars (ON) and emission at
levels below the sensitivity of current instrumentation (OFF).  In
these three cases, the time spent in the two states is measured in
weeks or years, long enough for the changes in spin-down rate to be
determined during the ON phases and hence to be deduced in the OFF
phases: if the frequency slow-down rates for the two phases are
$\dot\nu_{\rm ON}$ and $\dot\nu_{\rm OFF}$, then the long-term
slow-down rate is expected to be given by
\begin{equation}
\dot\nu = \dot\nu_{\rm OFF}\times(1-f_{\rm ON}) + \dot\nu_{\rm ON}\times f_{\rm ON} 
  = \dot\nu_{\rm OFF} + (\dot\nu_{\rm ON} - \dot\nu_{\rm OFF})\times f_{\rm ON}.
\label{eq:slowdown}
\end{equation}
Measurement of the long-term value of $\dot\nu$ as well as $\dot\nu_{\rm ON}$
and $f_{\rm ON}$ thus also allows the determination of $\dot\nu_{\rm OFF}$:
\begin{equation}
\dot\nu_{\rm OFF}=(\dot\nu - \dot\nu_{\rm ON}\times f_{\rm ON})/(1-f_{\rm ON}).
\end{equation}

These measurements have been carried out for the three long-term
intermittent pulsars mentioned above and are presented in
Table~\ref{table:all_intermittent}.  The values of $\dot\nu_{\rm ON}$
are always significantly greater in magnitude than those of $\dot\nu_{\rm
OFF}$, so that their ratio is always more than unity. For one of
the pulsars, B1931+24 \citep{ysl+13}, and now for J1910+0517
(\S\ref{sec:J1910+0517}), the duty cycle $f_{\rm ON}$ is stable over
several years.

In the case of PSR~J1929+1357, we have a new situation, in which the
ON times are much too short for the direct measurement of the spin-down
rate in each ON phase.  However, for the first time, we have been able
to determine a statistical long-term variation in the duty cycle
$f_{\rm ON}$, the fraction of time spent in the ON phase, and to
measure a corresponding change in the slow-down rate.  We note that such
an evolution in the statistical properties of the switching of
emission mode is similar to that demonstrated in PSRs~B1822$-$09 and
B1828$-$11 \citep{lhk+10}. In those cases, the switching is between
states with two different emission profiles and slow-down rates.

\begin{figure*}[p]
\begin{center}
\includegraphics[width=5.0in,angle=0]{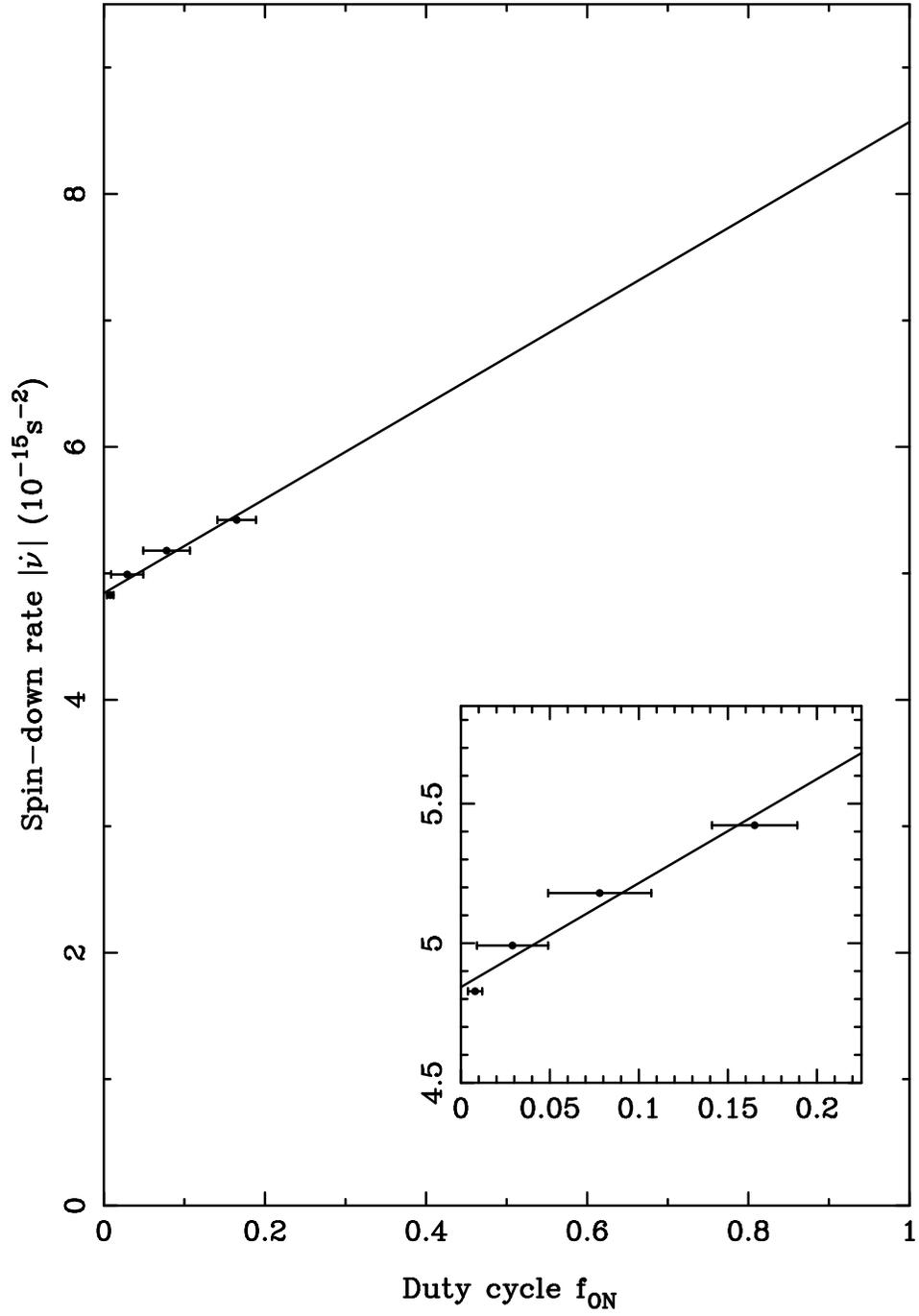}
\caption{\label{fig:1929_duty_f1} A plot of the magnitude of the
rotational spin-down rate $|\dot\nu|$ of PSR~J1929+1357 plotted
against the duty cycle $f_{\rm ON}$, which shows how the spin-down rate
depends upon the amount of radio emission from the pulsar.  }
\end{center}
\end{figure*}

The linear fits for the duty cycle given in Figure
\ref{fig:1929_stats}d are an approximation to the true form of the
variation, and we have performed a series of straight-line fits over
somewhat shorter time intervals to obtain local values of the duty
cycle $f_{\rm ON}$.  Over the same timespan of each of these fits, the
value of the first rotational frequency derivative $\dot\nu$ is
obtained from the fitted ephemeris, and the results are presented in
Figure \ref{fig:1929_duty_f1}. There is a clear increase in the
magnitude of the spin-down rate as the duty cycle of the intermittency
increases.

A least-squares fit of Equation \ref{eq:slowdown} to the data gives
values of $\dot\nu_{\rm OFF}=-4.84(3)\times10^{-15}$ s$^{-2}$ and
$\dot\nu_{\rm ON}=-8.6(5)\times10^{-15}$ s$^{-2}$, so that
$\dot\nu_{\rm ON}/\dot\nu_{\rm OFF}=1.8(1)$.

This dependency of the slow-down rate upon the emission state is
attributed to variations in the magnetospheric properties of the
pulsar, such as changes in plasma currents, which modify both the
emission from the magnetosphere and also the rotational torque and
hence the rate of loss of angular momentum \citep{klo+06}.  In this
pulsar, particles are responsible for increasing the spin-down rate by
about 80\%, comparable with the values obtained for the other 3
long-term intermittent pulsars presented in
Table~\ref{table:all_intermittent}. 

\section{Discussion}\label{sec:discussion}

The discovery and statistical study of the long-term intermittent
pulsars reported here reveal a large underlying population of such
ephemeral objects.  As pointed out by \citet{klo+06}, in single-pass
surveys, such as the PALFA survey, only a fraction $f_{\rm ON}$ of these
pulsars are detected during the survey observations.  Moreover,
even for the few which are detected, it is difficult to assess the
probability that such candidates will be confirmed as pulsars, because
confirmation strategies differ between surveys and usually limit the
number of reobservations $n_{\rm REOBS}$ to a handful because of 
constraints on telescope time.  As a result, in an extreme case like
PSR~J1929+1357, the probability of confirming a discovery is approximately 
$f_{ON}\times n_{\rm REOBS}$, so that the probability of making a confirmed
discovery is $f_{\rm ON}^{2}\times n_{\rm REOBS}$.  In 2013, with $f_{\rm ON}=0.008$ and
a reasonable value of $n_{\rm REOBS}=5$, the probability of a making a
confirmed discovery of a pulsar was $3\times10^{-4}$.  We note that,
ideally, reobservations should also be made on a specifically designed
set of timescales in order to capture pulsars with the wide range of
switching timescales that we now see to be possible in this class of
objects.

In the case of PSR~J1929+1357, it was fortunate that the pulsar was ON
during the second attempted reobservation; in the next 93 observations, it
was OFF.  Had the first 5 reobservations been unsuccessful, it is
possible that even such a strong pulsar would still remain unconfirmed.
Similar pulsars, but with one tenth of the flux density, must surely
exist and would be clearly detected in the survey observations,
particularly with the high sensitivity of the Arecibo Telescope used
for the PALFA survey, but the case for extended reobservation efforts
would be even less compelling: the unconfirmed detection would be
attributed to radio-frequency interference or to the detection of a normal
pulsar in a distant sidelobe.

For these reasons, there may be as many as several thousand 
strong pulsars like PSR~J1929+1357 in the sky, which happen to be OFF
when surveyed or, if ON, are OFF during the confirmation observations.
Similar but less severe selection effects preventing the discovery
of objects like the other long-term intermittent pulsars also
contribute to what must be a population of neutron stars of
significant size compared to that of normal pulsars.  As we noted earlier,
RRATs also suffer similar selection effects in one-pass surveys.  It
is likely that UTMOST \citep{cfb+16} and upcoming new projects, like
CHIME \citep{baa+14} and MeerTRAP (Stappers, B. W. et al. 2016, in
preparation) on
MeerKAT, will be useful for finding more such objects as they will make
many searches of the same piece of sky on various timescales for
single-pulse and periodic sources.

The discovery of two new long-term intermittent pulsars,
PSRs~J1910+0517 and J1929+1357, significantly expands the known
population of such objects, whose properties are summarized in
Table~\ref{table:all_intermittent}.  Although their number is still
small (only five such pulsars), we are now in a position to make a few
general remarks regarding their properties.

\begin{figure*}[p]
\begin{center}
\includegraphics[width=4.5in,angle=-90]{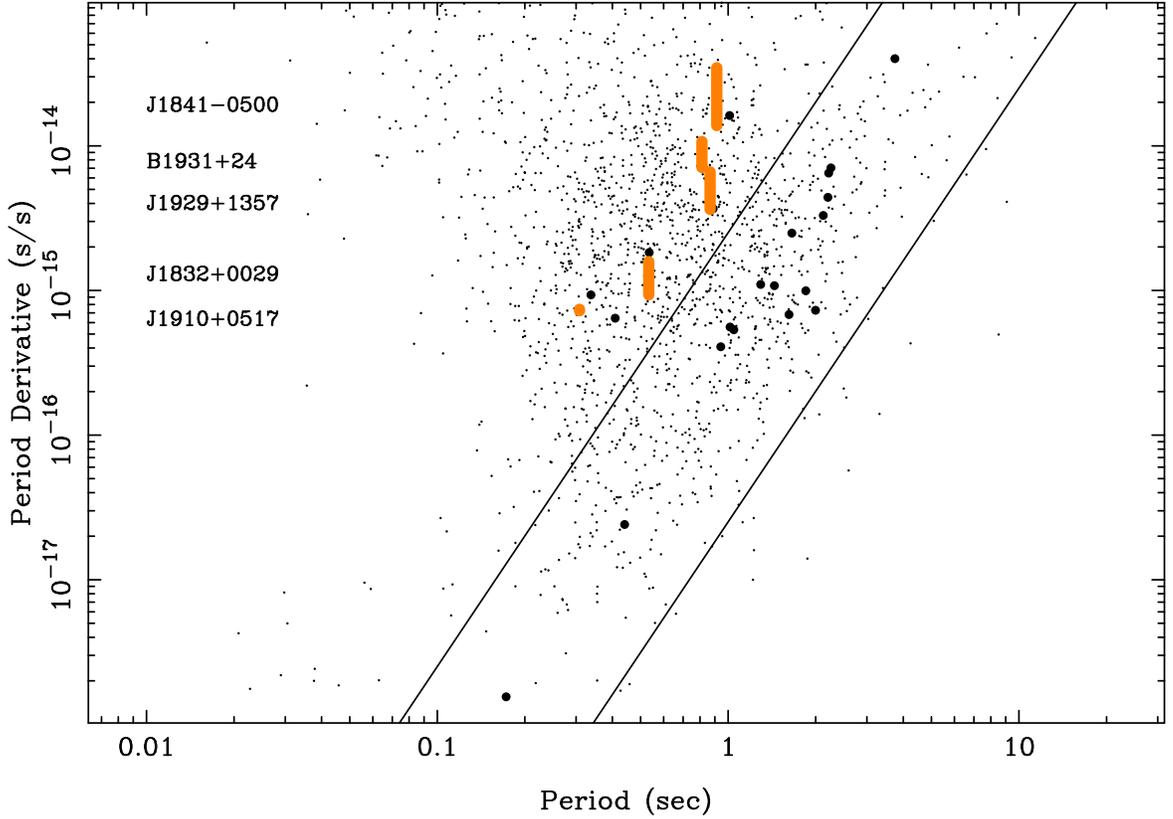}
\caption{\label{fig:ppdot} A plot of the rotational period derivative
$\dot P$ of pulsars plotted against the period $P$.  The vertical
orange lines represent the changes in positions of the five long-term
intermittent pulsars listed in Table~\ref{table:all_intermittent}
between the ON (top) and OFF (bottom) states.  The large black symbols
are at the positions of those pulsars which have published values of
null fractions \citep{wmj07} of greater than 15\% ($f_{\rm
ON}<85\%$). The sloping lines are the lines along which the rates of
loss of rotational kinetic energy are $\dot E=10^{32}$ erg s$^{-1}$
(upper) and $\dot E=10^{30}$ erg s$^{-1}$ (lower). These lines are
also parallel to lines of constant accelerating potential above the
polar cap. }
\end{center}
\end{figure*}

They seem to have a rather narrow range of spin periods, from about
0.3 to 0.9 s. Furthermore, there seems to be a positive correlation
between the spin period and its derivative, as we can see from their
ON positions in the $P$-$\dot{P}$ diagram (Figure~\ref{fig:ppdot}).
This correlation is similar to that observed for the shorter-term
nulling pulsars, although for the same $\dot{P}$ the latter have spin periods
$\sim\, 2.5$ times as long as those of the former.

In both cases, lines of constant $\dot{E}$ give a rough approximation
of the correlation between $P$ and $\dot{P}$, with $\dot{E}\, \sim \,
5 \, \times \, 10^{32}\, \rm erg \, s^{-1}$ representing a good fit
for the intermittent pulsars and $\dot{E}\, \sim \, 1 \, \times \,
10^{31}\, \rm erg \, s^{-1}$ being for the short-term nulling pulsars
(although the latter has a few significant outliers).  We note that 
the lines of constant $\dot{E}$ also have the same functional
dependence on $P$ and $\dot{P}$ ($\dot{P}/P^3$ = constant) as lines of
constant accelerating potential above the polar cap.  Since the
phenomenon we are discussing involves the cessation of plasma flow,
the accelerating potential may be a relevant parameter.
No other
quantity (e.g. age or B-field) appears to give such a good approximation of
the correlation between $P$ and $\dot{P}$ for these objects.
However, with such small-number statistics, at least for the long-term 
intermittent pulsars, any conclusions regarding this correlation must
be regarded as tentative. These relations offer new input for the
variety of theoretical models to explain their behavior, However, 
more discoveries of intermittent pulsars are
necessary for confirming (or refuting) this conclusion.

Conventional understanding of Figure~\ref{fig:ppdot} is that pulsars are formed
in the upper left and move down and across to the right.  Because of
the logarithmic form of the diagram, if pulsars maintain their
brightness throughout their lives, their density in this diagram
should increase by a factor of 10 every semi-decade.
Clearly this is not happening, and the density peaks at around 0.5 s
and then falls rapidly. The flux density and hence the luminosity of
pulsars must fall to take them below the sensitivity of our surveys.
This decrease may be continuous, as the rotation rate decreases and
the electrodynamic particle acceleration processes in the pulsar
magnetosphere reduce, or it is possible that it occurs in a stuttering
manner as the pulsar starts nulling, and an increase in the nulling
fraction eventually brings it permanently to the OFF state. The
presence of these nulling/intermittent pulsars in this unexpectedly
sparse region of the diagram suggests that the latter may occur
as the rate of loss of kinetic energy $\dot{E}$ reduces.  What is not
clear is why the long-term intermittent pulsars lie at values
of $\dot{E}$ higher than those where shorter-term nulling pulsars lie.

\acknowledgements 

The Arecibo Observatory is operated by SRI International under a
cooperative agreement with the National Science Foundation
(AST-1100968) and in alliance with Ana G. M\'endez-Universidad
Metropolitana and the Universities Space Research Association.
Our pulsar research at Jodrell Bank and access to the Lovell Telescope is
supported by a Consolidated Grant from the UK's Science and Technology
Facilities Council. This work was supported by the Max Planck
Gesellschaft and by NSF grants 1104902, 1105572, and 1148523. PCCF, PL,
and LGS gratefully acknowledge financial support from the European
Research Council for the ERC Starting Grant BEACON under contract
no. 279702. JvL acknowledges funding from the European Research
Council under the European Union's Seventh Framework Programme
(FP/2007-2013) / ERC Grant Agreement No. 617199. JSD was supported
by the NASA Fermi Guest Investigator program and by the Chief of Naval
Research.  JWTH acknowledges funding from an NWO Vidi fellowship
and from the European Research Council under the European Union's
Seventh Framework Programme (FP/2007-2013) / ERC Starting Grant
Agreement No. 337062 ("DRAGNET").  Pulsar research at UBC is supported
by an NSERC Discovery Grant and by the Canadian Institute for Advanced
Research.  VMK receives support from an NSERC Discovery Grant, an
Accelerator Supplement; a Gerhard Herzberg Award, and an
R. Howard Webster Foundation Fellowship from the Canadian Institute
for Advanced Study, the Canada Research Chairs Program, and the Lorne
Trottier Chair in Astrophysics and Cosmology. The National Radio Astronomy
Observatory is a facility of the National Science Foundation operated
under cooperative agreement by Associated Universities, Inc..

%\bibliography{journals,modrefs,psrrefs,crossrefs}
%\bibliographystyle{apj}

\begin{deluxetable}{lcc}
\tabletypesize{\scriptsize}
\tablewidth{0pc}
\tablecaption{Observed and Derived Parameters of intermittent PSRs~J1910+0517 and J1929+1357 \tablenotemark{a}\label{table:intermittent}}
\tablehead{                               & PSR~J1910+0517                        & PSR~J1929+1357}
\startdata
Right Ascension (J2000)                   & $19^h 10^m 37^s\!.907(14)$   & $19^h 29^m 10^s\!.62(2)$      \\
Declination (J2000)                       & $+5^\circ 17'56''\!.1(5)$    & $+13^\circ 57' 35''\!.9(5)$ \\
Galactic longitude                        & 38$^\circ$.84                & 49$^\circ$.63 \\ 
Galactic latitude                         & $-$1$^\circ$.83              & $-$1$^\circ$.81 \\
Rotation frequency (s$^{-1}$)             & 3.24624790060(12)            & 1.15350009753(9)              \\
Frequency first derivative (s$^{-2}$)     & $-7.698(8) \times 10^{-15}$  & $-4.87(3) \times 10^{-15}$ \\
Frequency second derivative (s$^{-3}$)    & $7.5(6) \times 10^{-24}$     & $10(2) \times 10^{-24}$ \\
Frequency third derivative (s$^{-4}$)     & $-0.18(3) \times 10^{-30}$   & $-0.42(4) \times 10^{-30}$ \\
Frequency fourth derivative (s$^{-5}$)    & $-2.6(3) \times  10^{-38}$   & $-$ \\
Epoch of pulsar frequency (MJD)           & 56700                        & 56440                      \\                   
Data span (MJD)                           & 55894$-$57400                & 56331$-$57407      \\ 
Half-power Pulse width $W_{50}$ (ms)      & 13.0                         & 11.5                   \\
Dispersion Measure DM (cm$^{-3}$pc)       & 300(2)                       & 150.7(3)         \\ 
Distance $d$\tablenotemark{b} (kpc)       & 7.3                          & 5.3 \\
\\
Mean Flux density ON S$_{\rm ON}$ (mJy)   & 0.5(1)                       & 2.2(4)                   \\ 
Mean Flux density OFF S$_{\rm OFF}$ (mJy) & 0.010(5)                     & 0.002(5)                  \\ 
Radio Luminosity ON L$_{\rm ON}=$S$_{\rm ON}d^2$ (mJy kpc$^2$)    &27  &63 \\
Radio Luminosity OFF L$_{\rm OFF}=$S$_{\rm OFF}d^2$ (mJy kpc$^2$)&0.9 &$<0.2$ \\
Radio Luminosity ratio L$_{\rm OFF}$/L$_{\rm ON}$                 &0.02 &$<0.003$ \\
Fraction of time in active mode $f_{\rm ON}$ & 0.30(4)                   & 0.008--0.165 \\
Duration of active modes, T$_{\rm ON}$ (h) & 6(3)                        & 2(1)--1.0(3) \\
Mean activity cycle time, T$_{\rm CYCLE}$ (h)& 19(8)                     & 220(110)--6.3(1.6) \\
Frequency derivative ON $\dot\nu_{ON}$ ($s^{-2}$) & $-$ & $-8.6(5)\times10^{-15}$ \\                      
Frequency derivative OFF $\dot\nu_{ON}$ ($s^{-2}$) & $-$ & $-4.84(3)\times10^{-15}$ \\                      
Spin-down age (Myr)                       & 6.68                         & 3.75 \\
Spin-down luminosity ($\mbox{erg/s}/10^{32}$) & 1.4                          & 2.2 \\
Inferred Magnetic Field (G$/10^{12}$)     & 3.0                          & 1.8 \\
\enddata
\tablenotetext{a}{Figures in parentheses are uncertainties in the last digit quoted.}
\tablenotetext{b}{Values predicted based on $l$, $b$, and DM, using the NE2001 electron density model of \cite{cl02}.}
\end{deluxetable}

\begin{deluxetable}{lcccccccl}
\tabletypesize{\scriptsize}
\tablewidth{0pc}
\tablecaption{Long-term intermittent pulsars \label{table:all_intermittent}}
\tablehead{ Pulsar & $\nu$ & $\dot\nu_{\rm OFF}$ & $\dot{E}$ & $\tau$ & $f_{\rm ON}$ & T & $\dot\nu_{\rm ON}/\dot\nu_{\rm OFF}$ &Reference\\
 & (Hz) & ($10^{-15}{\rm s}^{-2}$) & ($10^{32}{\rm erg\,s}^{-1}$) & (Myr) & & (d) & & }
\startdata
J1832+0029   & 1.873 & --3.3  & 2.4 &  8.9 & 0.6       & 2000 & 1.7(1) & \citet{llm+12}  \\
J1841$-$0500 & 1.095 & --16.7 & 7.2 &  1.0 & 0.5       & 800  & 2.5(2) & \citet{crc+12}  \\
J1910+0517   & 3.246 & --7.7  & 9.9 &  6.7 & 0.3       & 1    & --     & This paper  \\
J1929+1357   & 1.153 & --4.8  & 2.2 & 15.8 & 0.01-0.17 & 1-10 & 1.8(1) & This paper  \\
B1931+24     & 1.229 & --10.8 & 5.2 &  1.8 & 0.2       & 40   & 1.5(1) & \citet{klo+06}  \\
\enddata
\end{deluxetable}

\end{document}